# Structural Features of Layered Iron Pnictide Oxides $(Fe_2As_2)(Sr_4M_2O_6)$

H. Ogino[1,2,*], S. Sato[1], Y. Matsumura[1], N. Kawaguchi[1], K. Ushiyama[1], Y. Katsura[3], S. Horii[4], K. Kishio[1,2] and J. Shimoyama[1,2]

1 *Department of Applied Chemistry, University of Tokyo, 7-3-1 Hongo, Bunkyo-ku, Tokyo 113-8656, Japan*

2 *JST-TRIP, Sanban-cho, Chiyoda-ku, Tokyo 102-0075, Japan*

[3]*Magnetic Materials Laboratory, RIKEN, 2-1 Hirosawa, Wako-shi, Saitama 351-0198, Japan*

4 *Kochi University of Technology, Kami, Kochi 782-8502, Japan*

**Abstract**

Structural features of newly found perovskite-based iron pnictide oxide system have been systematically studied. Compared to *RE*FePnO system, perovskite-based system tend to have lower *Pn*-Fe-*Pn* angle and higher pnictogen height owing to low electronegativity of alkaline earth metal and small repulsive force between pnictogen and oxygen atoms. As-Fe-As angles of $(Fe_2As_2)(Sr_4Cr_2O_6)$, $(Fe_2As_2)(Sr_4V_2O_6)$ and $(Fe_2Pn_2)(Sr_4MgTiO_6)$ are close to ideal tetrahedron and those pnictogen heights of about 1.40 Å are close to NdFeAsO with optimized carrier concentration. These structural features of this system may leads to realization of high $T_c$ superconductivity.

*Keyword*: perovskite structure, iron pnictide, superconductor

---

[*] Corresponding author. Tel.: +81-3-5841-7777; fax: +81-3-5689-0574; e-mail: tuogino@mail.ecc.u-tokyo.ac.jp (H. Ogino).

## 1. Introduction

Since the discovery of high-$T_c$ superconductivity in $RE$FeAsO($RE$ = rare earth) system[1], several groups of iron pnictides or chalcogenides are developed. Meanwhile, still there are continuous demands for new materials containing iron tetragonal lattice. Recently new group of iron pnictides were discovered which consist of stacking of Fe$Pn$ ($Pn$ = pnictides) layers and perovskite-type oxide layers[2-7]. Superconductivity at 17 K in (Fe$_2$P$_2$)(Sr$_4$Sc$_2$O$_6$), 39 K in (Fe$_2$As$_2$)(Sr$_4$(Mg,Ti)$_2$O$_6$) and 46 K at 4 GPa in (Fe$_2$As$_2$)(Sr$_4$V$_2$O$_6$) show potential of high $T_c$ superconductivity in this system. From the view point of local structure, high symmetry in Fe$Pn_4$ tetrahedra[8] or high pnictogen height[9] are pointed out to be important factors for high $T_c$ superconductivity. From these insights, in this study we performed structural refinement of newly found (Fe$_2$As$_2$)(Sr$_4$MgTiO$_6$) and investigated structural features of the perovskite-based iron pnictide system.

## 2. Experimental

All samples were synthesized by the solid-state reaction starting from FeAs, SrO, Mg, Ti, TiO$_2$ and so on. Details of the sample preparation are found elsewhere[4,7]. Constituent phases and lattice constants were analyzed by the powder X-ray diffraction (XRD) method. Structural refinement was performed using the analysis program RIETAN-2000[10].

## 3. Results and discussion

Figure 1 shows the Rietveld refinement of the powder XRD pattern of (Fe$_2$As$_2$)(Sr$_4$MgTiO$_6$). Although the compound was formed as main phase, there were small amount of impurities such as SrFe$_2$As$_2$ and Sr$_2$TiO$_4$. Space group of the compound is *P4/nmm* and lattice constants were $a$ = 3.935 Å and $c$ = 15.95 Å, respectively. Structural parameters for (Fe$_2$As$_2$)(Sr$_4$MgTiO$_6$) are listed in Table 1.

In order to investigate structural features of the perovskite-based iron pnictide oxides, pnictogen heights and $\alpha$ angles of Fe$Pn$ layer of reported phases are arranged by their *a*-axis length as shown in Fig. 2. Both pnictogen heights and $\alpha$ angles depend on *a*-axis length while there are differences in trends derived from structural differences. Pnictogen heights of the perovskite-based systems are higher than that of $RE$FeAsO system. Possible reason is that the difference of neighboring atom to pnictogen site. Because alkaline earth element has lower electronegativity than rare earth atom, it attract pnictogen atom more than rare earth atom. Another explanation is small repulsive force between pnictogen and oxygen atom, because in perovskite-based system these atoms are screened by the cations. $Pn$-Fe-$Pn$ angles of (Fe$_2$As$_2$)(Sr$_4$Cr$_2$O$_6$), (Fe$_2$As$_2$)(Sr$_4$V$_2$O$_6$) and (Fe$_2$As$_2$)(Sr$_4$MgTiO$_6$) are close to the angle of ideal Fe$Pn_4$ tetrahedron (109.5°). Those pnictogen heights of about 1.4 Å are close to NdFeAs(O,F) with optimized carrier concentration as well as FeSe under pressure. Those structural features of perovskite-based iron pnictide oxides are ideal from the viewpoints of both Lee and Kuroki pointed.

## 4. Conclusions

In the present study, structural features of perovskite-based iron pnictide oxides have been strudied. Perovskite-based system has higher pnictogen height and lower $Pn$-Fe-$Pn$ angle compared to those of $RE$FeAsO. These structural features of this system may leads to higher $T_c$.

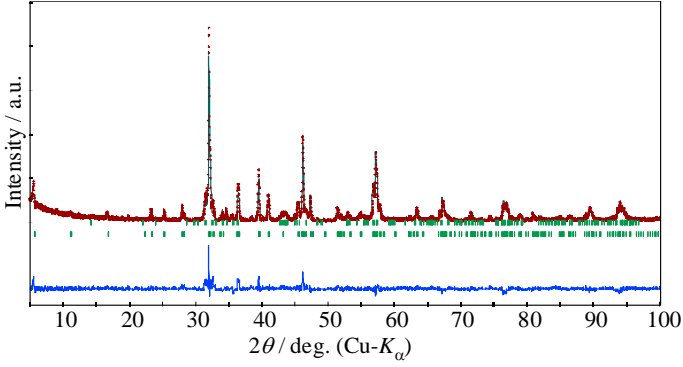

Fig. 1 Rietveld refinement of powder XRD pattern of $(Fe_2As_2)(Sr_4MgTiO_6)$: dashed line indicates calculated fit pattern and solid line indicate difference of observed and calculated pattern. Upper and lower bars show diffraction peak positions of the $SrFe_2As_2$ and $(Fe_2As_2)(Sr_4MgTiO_6)$, respectively.

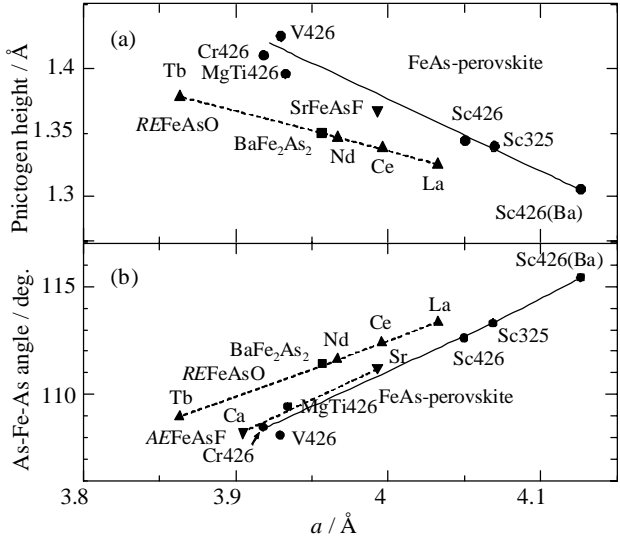

Fig. 2 Relationship between pnictogen height and $a$ axis length (a) and As-Fe-As angle and $a$ axis length (b) of several iron arsenides. $(Fe_2As_2)(Sr_3M_2O_5)$, $(Fe_2As_2)(Sr_4M_2O_6)$ and $(Fe_2As_2)(Ba_4M_2O_6)$ are abbreviated as $M325$, $M426$ and $M426(Ba)$, respectively. Data are replotted from ref[4].

Table. 1 Structural parameters for $(Fe_2As_2)(Sr_4MgTiO_6)$

| Atom | $x$ | $y$ | $z$ |
|---|---|---|---|
| Mg/Ti | 0.250 | 0.250 | 0.3086 |
| Fe | 0.250 | -0.250 | 0.0000 |
| Sr1 | -0.250 | -0.250 | 0.1900 |
| Sr2 | -0.250 | -0.250 | 0.4133 |
| O1 | 0.250 | -0.250 | 0.2911 |
| O2 | 0.250 | 0.250 | 0.4275 |
| As | 0.250 | 0.250 | 0.0864 |

$R_{wp}$ = 16.08 $R_p$ = 12.06 $R_e$ = 18.60 $S$ = 0.8647 $R_F$ = 3.67